\documentstyle[sprocl]{article}

\def\be{\begin{equation}}
\def\ee{\end{equation}}
\def\bea{\begin{eqnarray}}
\def\eea{\end{eqnarray}}

\begin{document}

\title{The Semiclassical Approach to Small $x$ Physics\footnote{Talk 
given at the DIS-2000 conference, Liverpool, England, April 2000}}

\author{{\bf \it {A. Metz}}}

\address{CEA-Saclay, DAPNIA/SPhN, F-91191 Gif-sur-Yvette, France}

\author{{\bf \it {H.G. Dosch, A. Hebecker, H.J. Pirner}}}

\address{Institut f\"ur Theoretische Physik, Universit\"at Heidelberg,
Philosophenweg 16 $\&$ 19, D-69120 Heidelberg, Germany}

\maketitle

\abstracts{The semiclassical approach to small-x physics is briefly 
reviewed, where the main emphasis is on the determination of the gluon 
distribution at NLO.}

\section{Introduction}
The semiclassical approach to deep inelastic scattering (DIS) at small
$x$ exploits the target rest frame point of view \cite{buchmueller_96}.
In this frame the virtual photon interacts via its partonic fluctuation 
with the target.
The target itself is considered as localized soft color field.
In the small $x$ limit the partons are fast, and therefore their
interaction with the target can be treated in an eikonal 
approximation \cite{nachtmann_91}.

This picture of DIS allows a combined description of both inclusive 
and diffractive events.
In particular in the case of diffractive scattering, like diffractive 
dissociation and the diffractive production of high-$p_{\perp}$ jets,
several interesting results can be obtained, even without explict 
numerical calculations.\footnote{For a comprehensive review of these 
topics see Ref. \cite{hebecker_99} and references therein.}

Here we focus on the most recent developement in the semiclassical 
approach, namely the determination of the gluon density at
next-to-leading order (NLO) \cite{dosch_00}, which can serve as input in 
the evolution equation.
The gluon density is expressed in terms of a (non-perturbative) Wilson
loop and can be evaluated in any model of the target color field. 

\section{The Gluon Distribution}
To extract the gluon density it is convenient to use a `scalar 
photon' (denoted by $\chi$) coupled directly to the gluon 
field \cite{mueller_90}. 
The gluon density is then derived by matching the semiclassical and the 
parton model approach.
To leading order this means that we have to equate the cross section for 
the transition $\chi \to g$ in an external field with the cross section 
of the process $\chi g \to g$ as given in the parton model.
The result reads
\begin{equation} \label{gd_lo}
xg^{(0)}(x,\mu^2) = \frac{1}{12\pi^2\alpha_s} 
\int d^2x_{\perp} \left| \frac{\partial}{\partial y_\perp}
W^{\cal A}_{x_\perp}(y_\perp) \Big|_{y_\perp=0} \right|^2 \,,
\end{equation}
where $W^{\cal A}$ indicates a Wilson loop in the adjoint 
representation describing the eikonalised interaction of a gluon 
in the external color field of the target.
The gluon distribution $xg^{(0)}(x,\mu^2)$ is a constant, and measures 
the averaged local field strength of the target.

At NLO, we write the gluon density as
\begin{equation} \label{gd_sum}
xg(x,\mu^2) = xg^{(0)}(x,\mu^2) + xg^{(1)}(x,\mu^2)\,,
\end{equation}
with $xg^{(1)}(x,\mu^2)$ denoting the (scheme dependent) NLO 
correction.
To extract this correction, the cross section for the transition 
$\chi \to gg$ in an external field has to be equated with the parton model
cross section of the process $\chi g \to gg$.
Without providing any details of the calculation, we quote here only the 
final result of the distribution in the $\overline{\mbox{MS}}$ scheme
at NLO \cite{dosch_00},  
\begin{equation} \label{gd_nlo}
xg^{(1)}(x,\mu^2) = \frac{1}{\pi^3} \left( \ln\frac{1}{x} \right) \,
\int_{r^2(\mu)}^{\infty} \, \frac{dy_\perp^2}{y_\perp^4}
\left\{ -\int d^2x{_\perp} \,\mbox{tr} \, 
W^{\cal A}_{x_\perp}(y_{\perp}) \right\}\,.
\end{equation}
The scheme dependence enters through the short-distance cutoff 
$r^{2}(\mu)$.
The NLO gluon density shows a $\ln (1/x)$ enhancement at small $x$, 
and is sensitive to the large-distance structure of the target.
\\
Using the model of a large hadron to describe the color field of the 
target, allows a comparison of our result with the one of 
Mueller \cite{mueller_90,mueller_99}.
We find agreement for both the integrated distribution in (\ref{gd_nlo}) 
and the unintegrated one \cite{dosch_00} not shown here.
However, in Refs. \cite{mueller_90,mueller_99}, where the main focus is
on parton saturation, the scale dependence has not been discussed.
Therefore, we provide for the first time a quantitative relation
between the short-distance cutoff in Eq. (\ref{gd_nlo}) 
and the scale of the gluon distribution, which can only be achieved by 
matching the semiclassical approach with a treatment in the parton model.



\section*{References}
\begin{thebibliography}{99}
\bibitem{buchmueller_96}
 W. Buchm\"uller and A. Hebecker, {\it Nucl. Phys.} B {\bf 476}, 
 203 (1996). 
\bibitem{nachtmann_91}
 O. Nachtmann, {\it Ann. Phys.} {\bf 209}, 436 (1991).
\bibitem{hebecker_99}
 A. Hebecker, {\it Phys. Rept.} {\bf 331}, 1 (2000).
\bibitem{dosch_00}
 H.G. Dosch, A. Hebecker, A. Metz and H.J. Pirner, {\it Nucl. Phys.}
 B {\bf 568}, 287 (2000).
\bibitem{mueller_90}
 A.H. Mueller, {\it Nucl. Phys.} B {\bf 335}, 115 (1990).
\bibitem{mueller_99}
 A.H. Mueller, {\it Nucl. Phys.} B {\bf 558}, 285 (1999).

\end{thebibliography}
\end{document}